\documentclass{mecbic}
\providecommand{\anno}{2011}
\providecommand{\firstpage}{83}
\providecommand{\lastpage}{94}
\usepackage{breakurl}        

\usepackage[ansinew]{inputenc}
\usepackage{amssymb}
\usepackage{amsmath}
\usepackage{amsthm}
\usepackage{xspace}
\usepackage{url}
\usepackage{color}
\usepackage{graphicx}
\usepackage{subfigure}
\usepackage[english]{babel}
\usepackage{chemarrow}

\graphicspath{{Fig/}}

\newtheorem{theorem}{Theorem}

\theoremstyle{definition}
\newtheorem{definition}{Definition}

\newcommand{\syn}[6]{\hbox{\ensuremath{(#1,#2)(#4,#5)\to(#1,#3)(#4,#6)}}}
\newcommand{\token}{\ensuremath{\bullet}}
\newcommand{\synt}[4]{\hbox{\ensuremath{(\token,#1)(\token,#2)\to(\token,#3)(\token,#4)}}}
\newcommand{\syntp}[5]{\hbox{\ensuremath{(\token,#1)(\token,#2)\to^{#5}(\token,#3)(\token,#4)}}}
\newcommand{\To}{\Rightarrow}
\newcommand{\Top}[3]{\hbox{\ensuremath{#1\To^{#3}#2}}}

\newlength{\fleche}
\newcommand{\mfd}[1]{\ensuremath{\ \autorightarrow{\ensuremath{#1}}{}{}\ }}

\newcommand{\mgs}{\textsf{MGS}\xspace}

\title{Generalized Communicating P Systems\\Working in Fair Sequential Mode}
\author{
Antoine Spicher
\institute{LACL, D\'epartement Informatique\\ Universit\'e Paris Est\\ 61 av. g\'en. de Gaulle, 94010, Cr\'eteil, France}
\email{antoine.spicher@u-pec.fr} \and
Sergey Verlan
\institute{LACL, D\'epartement Informatique\\ Universit\'e Paris Est\\ 61 av. g\'en. de Gaulle, 94010, Cr\'eteil, France}
\email{verlan@u-pec.fr}}
\def\titlerunning{Generalized Communicating P Systems Working in Fair Sequential Mode}
\def\authorrunning{Antoine Spicher and Sergei Verlan}

\begin{document}
\maketitle
\pagestyle{plain}
\pagenumbering{arabic}
\setcounter{page}{83}

\begin{abstract}
In this article we consider a new derivation mode for generalized communicating
P systems (GCPS) corresponding to the functioning of population protocols (PP)
and based on the sequential derivation mode and a fairness condition. We show
that PP can be seen as a particular variant of GCPS. We also consider a
particular stochastic evolution satisfying the fairness condition and obtain
that it corresponds to the run of a Gillespie's SSA. This permits to further
describe the dynamics of GCPS by a system of ODEs when the population size
goes to the infinity.
\end{abstract}

\section{Introduction}

The notion of a {\it generalized communicating P system} was introduced in
\cite{VBGM08}, with the aim of providing a common generalization of various
purely communicating models in the framework of P systems.

A generalized communicating P system, or a {\it GCPS} for short, corresponds to
a hypergraph where each node is represented by a cell and each edge is
represented by a rule. Every cell contains a multiset of objects which -- by
communication rules -- may move between the cells.  The form of a {\it
communication rule} is $\syn{a}{i}{k}{b}{j}{l}$ where $a$ and $b$ are objects
and $i,j,k,l$ are labels identifying the input and the output cells. Such a rule
means that an object $a$ from cell~$i$ and an object $b$ from cell~$j$ move
synchronously to cell~$k$ and cell~$l$, respectively.  In this respect, the
model resembles the Petri Net formalism~\cite{PN} where tokens from various
input places come along together to fire a given transition and then fork out to
destination places, see~\cite{VBGM08,BGMV07} for more details.

Depending on the communication rules form, several {\it restrictions on
communication rules} (modulo symmetry) can be introduced. Due to the simplicity
of their rules, the generative power of such restricted systems is of particular
interest and it has been studied in detail.
In~\cite{VBGM08,CV11,CVV10,Phandbook} it was proved that eight of the possible
nine restricted variants (with respect to the form of rules) are able to
generate any recursively enumerable set of numbers; in the ninth case only
finite sets of singletons can be obtained. Furthermore, these systems even with
relatively small numbers of cells and simple underlying (hypergraph)
architectures are able to achieve this generative power.  In~\cite{CVV10} a
further restriction is introduced by considering that the alphabet of objects is
a singleton (like in Petri Nets) and it is shown that the computational
completeness can be achieved in four of the restricted variants.

Population protocols (PP) have been introduced in~\cite{AADFP04}
(see~\cite{AR07} for a survey) as a model of sensor networks consisting of very
limited mobile agents with no control over their own movement. A population
protocol corresponds to a collection of anonymous agents, modelled by finite
automata, that interact with one another to carry out computations, by updating
their states, using some rules. Their computational power has been investigated
under several hypotheses in most of the cases restricted to finite size
populations. In particular, predicates stably computable in the original model
have been characterized as those definable in Presburger arithmetic. The
article~\cite{Bournez} studies the convergence of PP when the population size
goes to the infinity.

The evolution of a PP follows a particular fairness condition: an execution is
\emph{fair} if for all configurations $C$ that appear infinitely often in the
execution, if $C$ is predecessor of a configuration~$C'$, then~$C'$ appears
infinitely often in the execution. We consider such a condition in the case of
GCPS systems and obtain a new derivation mode which we call \emph{fair
sequential mode} (fs-mode). We further study the dynamic behaviour of the system
in this mode. Among several possible evolution strategies we consider a
stochastic strategy that satisfies the fairness condition and we obtain that the
evolution of the system corresponds to a run of the Gillespie stochastic
simulation algorithm (SSA). Using the correspondence between SSA and ODEs
(assuming mass-action kinetics) we show that the dynamics of the system can be
represented by a system of ODEs when the population size goes to the
infinity. We also consider the converse problem and we give sufficient
conditions for a system of ODEs to be represented by a GCPS system working in
concentration-depended stochastic implementation of the fs-mode. We consider
several examples of GCPS simulating Lotka-Volterra (predator-prey) behaviour or
computing approximations of algebraic numbers.

\section{Background}

In this section we recall some basic notions and notations used in membrane
computing, formal language theory and computability theory. For further details
and information the reader is referred to~\cite{Pbook,Phandbook,HFL}.

An alphabet is a finite non-empty set of symbols. For an alphabet $V$, we denote
by $ V^*$ the set of all strings over $V$, including the empty string,
$\lambda$. The \emph{length} of the string $x\in V^*$ is the number of symbols
which appear in $x$ and it is denoted by $|x|$. The number of occurrences of a
symbol $a\in V$ in $x\in V^*$ is denoted by $|x|_a$. If $x\in{V^*}$ and
$U\subseteq V$, then we denote by $|x|_U$ the number of occurrences of symbols
from $U$ in $x$.

A finite multiset over $ V $ is a mapping $M: V \longrightarrow \mathbb{N} $;
$M(a)$ is said to be the multiplicity of $ a $ in $ M $ ($\mathbb{N}$ denotes
the set of non-negative integers.) A finite multiset $M$ over an alphabet $V$
can be represented by all permutations of a string $x=a_1^{M(a_1)}\,
a_2^{M(a_2)} \ldots a_n^{M(a_n)} \in V^* $, where $a_j \in V $, $ 1 \leq j \leq
n $; $x$ represents $M$ in~$V^*.$ If no confusion arises, we also may use the
customary set notation for denoting multisets. The size of a finite multiset
$M,$ represented by $x\in V^*$ is defined as $\Sigma_{a\in V} \vert x\vert_a$.

\subsection{P Systems}

Next we recall the basic definitions concerning generalized communicating P
systems ~\cite{VBGM08}.

\begin{definition}\label{def:gencom}
A \emph{generalized communicating P system} (a \emph{GCPS}) of degree $n,$
where $n\ge 1,$  is an $(n+4)$-tuple $ \Pi=(O,E,w_1,\dots{},w_n,R,h) $ where
\begin{enumerate}
\item $ O $ is an alphabet, called the \emph{set of objects} of $\Pi$;
\item $ E \subseteq O $; called the \emph{set of environmental objects} of
    $\Pi$;
\item $ w_i \in O^*$, $1 \leq i \leq n $, is the multiset of objects
    \emph{initially associated with cell $ i $};
\item $ R $ is a finite set of \emph{interaction rules} (or
    \emph{communication rules}) of the form \syn{a}{i}{k}{b}{j}{l}, where
    $a,b\in O$, $0\le i,j,k,l\le n$, and if
     $i=0$ and $j=0$, then $\{a,b\}\cap (O\setminus E)\ne \emptyset$; \emph{i.e.},
     $a\notin E$ and/or $b\notin E;$
\item $ h \in \{1,\ldots,n\}$  is the \emph{output cell}.
\end{enumerate}
\end{definition}

The system consists of $n$ cells, labelled by natural numbers from $1$ to $n$,
which contain multisets of objects over $O$; initially cell $i$ contains
multiset $w_i$ (the initial contents of cell $i$ is $w_i$). We distinguish an
additional special  cell, labelled by $0$, called the \emph{environment}. The
environment contains objects of $E$ in an \emph{infinite number of copies}.

The cells interact by means of the rules $\syn{a}{i}{k}{b}{j}{l}$, with $a,b\in
O$ and $0\le i,j,k,l\le n$. As the result of the application of the rule,
object $a$ moves from cell~$i$ to cell~$k$ and $b$ moves from cell~$j$ to
cell~$l$. If two objects from the environment move to some other cell or cells,
then at least one of them must not appear in the environment in an infinite
number of copies. Otherwise, an infinite number of objects can be imported in
the system in one step.

A \emph{configuration} of a GCPS $\Pi$, as above, is an $(n+1)$-tuple
$(z_0,z_1,\dots{},z_n)$ with $ z_0 \in (O \setminus E)^*$ and $ z_i \in O^*$,
for all $ 1 \leq i \leq n $; $ z_0 $ is the multiset of objects present in the
environment in a finite number of copies, whereas, for all $ 1 \leq i \leq n $,
$ z_i $ is the multiset of objects present inside cell $ i $. The \emph{initial
configuration of $ \Pi $} is the tuple $(\lambda,w_1,\dots{},w_n)$.

Given a multiset of rules $\cal{R}$ over $R$ and a configuration
$u=(z_0,z_1,\dots{},z_n)$ of $ \Pi$, we say that $\cal{R}$ is \emph{applicable}
to $u$ if all its elements can be applied simultaneously to the objects of
multisets $z_0, z_1,\dots{},z_n$ such that every object is used by at most one
rule.  Then, for a configuration $u=(z_0,z_1,\dots{},z_n)$ of $ \Pi$, a new
configuration $u'=(z'_0,z'_1,\dots{},z'_n)$ is obtained by applying the rules
of $R$ in a non-deterministic maximally parallel manner: taking an applicable
multiset of rules $\cal{R}$ over $R$ such that the application of~$\cal{R}$
results in configuration $u'=(z'_0,z'_1,\dots{},z'_n)$ and there is no other
applicable multiset of rules~$\cal{R'}$ over~$R$ which properly contains
$\cal{R}$.

It is also possible to replace the maximally parallel strategy of rule
application by other strategies, called \emph{derivation modes} (in the context
of the present paper, the terms \emph{mode} and \emph{strategy} are used
indifferently). A derivation mode lies in the heart of the semantics of P
systems and it permits to specify which multiset among different possible
applicable multisets of rules can be applied. When P systems were introduced,
only the maximally parallel derivation mode was considered which states that
corresponding multisets should be maximal, \emph{i.e.}, non-extensible.
With the appearance of the minimal parallel derivation mode~\cite{minpar} the
concept of the derivation mode had to be precisely defined
and~\cite{FreundVerlan07} presents a framework that permits to easily define
different derivation modes.

One application of a multiset of rules satisfying the conditions of a derivation
mode represents a \emph{transition} in $ \Pi $ from configuration $u$ to
configuration $u'$.  A transition sequence is said to be a {\it successful
generation} by $\Pi$ if it starts with the initial configuration of $\Pi$ and
ends with a \emph{halting configuration}, \emph{i.e.}, with a configuration
where no further transition step can be performed.

We say that $\Pi$ \emph{generates a non-negative integer} $n$ if there is a
successful generation by $\Pi$ such that $n$ is the size of the multiset of
objects present inside the output cell in the halting configuration.  The
\emph{set of non-negative integers generated} by a GCPS $ \Pi$ in this way is
denoted by $N(\Pi) $. It is also possible to use GCPS as acceptors, in this
case an input multiset is accepted if the system halts on it.

In~\cite{VBGM08} it is shown that GCPS are able to generate all recursively
enumerable languages. Moreover this result can be obtained by using various
restrictions on the type of rules (\emph{i.e.} induced hypergraph structures),
on the number of membranes and on the cardinality of the alphabet. We refer
to~\cite{VBGM08,CV11,CVV10} for more details.

If the cardinality of the alphabet $O$ is equal to one, then we refer to the
corresponding symbol as a token (denoted by $\token$). Hence, we assume that
$O=\{\token\}.$ We observe that such systems are similar to Petri Nets having a
restricted topology. This is especially visible if a graphical notation is
used.  However, the maximal parallelism and the concept of the environment are
specific to P systems, so we place this study in the latter framework.   A
converse study of P systems from the point of view of Petri Nets can be found
in~\cite{friscoBook}. For more details on Petri Nets and membrane computing we
also refer to \cite{Phandbook}.

In this article we shall consider the dynamics of the configuration of GCPS, so
we are no more interested in computation (and halting evolutions).

\subsection{Population Protocols}

We give below the definition as it appears in~\cite{Bournez}. A protocol is
given by $(Q,\Sigma, \i, \omega, \delta)$ with the following components. $Q$ is
a finite set of states. $\Sigma$ is a finite set of input symbols. $\i :
\Sigma\to Q$ is the initial state mapping, and $\omega : Q \to \{0, 1\}$ is the
individual output function. $\delta\subseteq Q^4$ is a joint transition relation
that describes how pairs of agents can interact. Relation $\delta$ is sometimes
described by listing all possible interactions using the notation
$(q_1,q_2)\to(q_1', q_2')$, or even the notation $q_1q_2\to q_1'q_2'$, for
$(q_1, q_2, q_1', q_2') \in\delta$ (with the convention that $(q_1,
q_2)\to(q_1,q_2)$ when no rule is specified with $(q_1, q_2)$ in the left hand
side.)

Computations of a protocol proceed in the following way. The computation takes
place among $n$ agents, where $n\ge 2$. A configuration of the system can be
described by a vector of all the agent's states. The state of each agent is an
element of $Q$. Because agents with the same states are indistinguishable, each
configuration can be summarized as an unordered multiset of states, and hence
of elements of $Q$.

Each agent is given initially some input value from $\Sigma$: each agent's
initial state is determined by applying $\i$ to its input value. This
determines the initial configuration of the population.

An execution of a protocol proceeds from the initial configuration by
interactions between pairs of agents. Suppose that two agents in state $q_1$ and
$q_2$ meet and have an interaction. They can change into state $q_1'$ and $q_2'$
if $(q_1, q_2, q_1', q_2')$ is in the transition relation $\delta$. If $C$ and
$C'$ are two configurations, we write $C\to C'$ if $C'$ can be obtained from $C$
by a single interaction of two agents: this means that $C$ contains two states
$q_1$ and $q_2$ and $C'$ is obtained by replacing $q_1$ and $q_2$ by $q_1'$ and
$q_2'$ in $C$, where $(q_1, q_2, q_1', q_2' )\in\delta$. An execution of the
protocol is a (potentially infinite) sequence of configurations
$C_0,C_1,C_2,\dots$, where $C_0$ is an initial configuration and
$C_i\to{C_{i+1}}$ for all $i\ge0$. An execution is \emph{fair} if for all
configurations $C$ that appears infinitely often in the execution, if $C\to C'$
for some configuration $C'$, then $C'$ appears infinitely often in the
execution.

At any point during an execution, each agent's state determines its output at
that time. If the agent is in state $q$, its output value is $\omega(q)$. The
configuration output is $0$ (respectively $1$) if all the individual outputs
are $0$ (respectively $1$). If the individual outputs are mixed 0s and 1s then
the output of the configuration is undefined.

Let $p$ be a predicate over multisets of elements of $\Sigma$. Predicate $p$
can be considered as a function whose range is $\{0, 1\}$ and whose domain is
the collection of these multisets. The predicate is said to be computed by the
protocol if, for every multiset $I$, and every fair execution that starts from
the initial configuration corresponding to $I$, the output value of every agent
eventually stabilizes to $p(I)$.

The following was proved in~\cite{AADFP04,AAE06}:

\begin{theorem}[\cite{AADFP04,AAE06}] A predicate is computable in the
population protocol model if and only if it is semilinear.
\end{theorem}

Recall that semilinear sets are known to correspond to predicates on counts of
input agents definable in first-order Presburger
arithmetic~\cite{Presburger29}.

\subsection{Gillespie Algorithm}

A usual abstraction in the simulation of biochemical systems consists in
considering the system (e.g., a bacterium) as a homogeneous chemical solution
where the reactions of the model are taking place. D.T. Gillespie has proposed
in~\cite{gillespie77} an algorithm for producing the trajectories of such a
chemical system by computing the \emph{next reaction} and the \emph{elapsed
time} since last reaction occurred. Let $\mu$ be a chemical reaction. The
probability that $\mu$ takes place during an infinitesimal time step is
proportional to:
\begin{itemize}
 \item $c_\mu$, the \emph{stochastic reaction constant}\footnote{Evaluating the
 stochastic constants is one of the key issues in stochastic simulations of
 biochemical reactions.} of reaction $\mu$;
 \item $h_\mu^S$, the number of distinct molecular combinations that can
 activate reaction $\mu$; it depends on the current chemical state $S$;
 \item $d\tau$, the length of the time interval.
\end{itemize}
Gillespie proved that the probability $P(\tau,\mu|S)d\tau$ that, being in a
chemical state $S$, the next reaction will be of type $\mu$ and will occur in
the time interval $(t+\tau,t+\tau+d\tau)$ is:
\begin{equation*}
 P(\tau, \mu|S)d\tau=a_\mu^S\,e^{-a_0^S\,\tau} d\tau 
\end{equation*}
where $a_\mu^S= c_\mu\,h_\mu^S$ is called the \emph{propensity} of reaction
$\mu$, and $a_0^S=\sum_\nu a_\nu^S$ is the combined propensity of all
reactions.

This probability leads to the first straightforward Gillespie's \emph{exact
stochastic simulation algorithm} (SSA) called the \emph{first reaction
method}. From a chemical state $S$, it consists in choosing an elapsed time
$\tau$ for each reaction $\mu$ according to the probability $P(\tau,\mu|S)$. The
reaction with the lowest elapsed time is selected and applied on the system
making its state evolve. A new probability distribution is then computed for
this new state and the process is iterated.

The Gillespie's SSA gives a way to simulate a continuous-time Markov chain with
the states corresponding to the states of the system and with transitions
between states corresponding to a single occurrence of a reaction. The
probability for a transition between two states $S$ and $S'$ corresponding to
the application of rule $\mu$ is defined as $a_\mu^S/a_0^S$.
In the following, we drop the mention of the current state $S$ in these
notations.

\section{Fair Sequential Derivation Mode}

In this section we are interested in the relation between PP and GCPS.
We show that in terms of structure PP and GCPS are quite similar, the main
differences concern the environment and the derivation mode. We define a new
\emph{fair sequential} mode for GCPS and hence we are able to encode any PP in a
GCPS w.r.t. their dynamics.
We then remark that GCPS with stochastic and Gillespie-like strategies are part
of this new class of GCPS and we propose their use for simulations of population
behaviours.

It can be easily seen that both PP and GCPS are particular instances of multiset
rewriting. Indeed, in both cases the underlying data structure is multiset
(obtained in a direct way for PP and by attaching the indices of membranes to
the objects in GCPS) and the evolution rules clearly correspond to multiset
rewriting rules with both left hand and right hand sides of size two.  So, the
translation of a PP to a one-symbol GCPS can be easily done as follows.  Given a
PP with set of states $Q$ (for convenience we suppose that $Q=\{1,\dots,n\}$) and
transition relation $\delta$ in an initial configuration $C_0$, the
corresponding GCPS $\Pi=(O,E,w_1,\dots{},w_n,R,1)$ can be defined as:
\begin{itemize}
 \item $O=E=\{\token\}$,
 \item $w_q = \token^k$, $k=|C_0|_q$ for any $q\in Q$,
 \item $R = \{\synt{q_1}{q_2}{q_1'}{q_2'}\;|\;q_1q_2\to q_1'q_2'\in\delta\}$.
\end{itemize}
The above system encodes each state $q$ of PP by a token $\token$ present in
membrane labelled by $q$. Since we are interested in the dynamics of the system,
no output membrane is necessary. The above construction covers the core of PP
and to obtain the complete equivalence encoding and decoding functions $\i$ and
$\omega$ shall be used in the same way.

It is not possible to do a similar encoding of GCPS with PP because PP always
deal with finite multisets and GCPS can use the infinite multiset corresponding
to the environment. However, any GCPS having no rule involving the environment
can be translated to PP in a similar way.

We remark that the biggest difference between PP and GCPS is given by the
evolution step, \emph{i.e.}, by the derivation mode. For GCPS, mainly the
maximally parallel derivation mode is investigated with several attempts to
investigate asynchronous or minimally parallel derivation mode,
see~\cite{Phandbook} for more details. The derivation mode of PP is very
particular -- it corresponds to a sequential strategy where only one rule is
applied at each step, like in Petri Nets, but with an additional fairness
condition.

We can consider such a strategy in GCPS case as well. More precisely we
consider
\emph{fair} computations: a GCPS computation is \emph{fair}, if for any
configuration $u$ that appears infinitely often in the computation, then any
configuration $u'$, such that $u\To{u'}$ in sequential application, also
appears infinitely often.
We shall call such computational strategy a \emph{fair sequential derivation
mode} (shortly \emph{fs-mode}).

From these considerations, it is trivial to observe that PP are similar to GCPS
in fs-mode with only one symbol in the alphabet. If we consider an encoding
function $\i$ like for PP and the halting condition corresponding to the
stabilization of the $\omega$-image of the configuration, then as an immediate
consequence of~\cite{AADFP04,AAE06}, we obtain that any GCPS working in fs-mode
and that does not have any rule involving the environment can only accept
semilinear sets.

Conversely, we also obtain that any PP working in maximally parallel mode
(\emph{i.e.}, a maximally parallel number of interactions can happen at each
step) are computationally complete if the number agents in some particular
state $q_0$ is going to the infinity.

From now on, we only speak of PP in terms of their associated one-symbol GCPS
in fs-mode.

\subsection{FS-Mode and Stochastic Evolution}

Although powerful the definition of the fairness remains obscure. Let try to
clarify it.
When assuming that the number of configurations is finite (this is the case for
classical PP for example), the definition can be easily rephrased as follows: a
computation $u_0\To{u_1}\To\dots$ is fair if
\begin{itemize}
 \item there exists a non-negative integer $N$ such that configuration $u_N$
 belongs to a terminal strongly connected component of the state
 graph\footnote{In this directed graph, nodes correspond to the configurations
 $u$, and two nodes $u$ and $u'$ are directly linked if $u\To{u'}$.}; and
 \item any state of this terminal strongly connected component appears
 infinitely often in the execution.
\end{itemize}
There are many possible evolutions of the system in the fs-mode. One example of
such an evolution is to choose at each step a rule that leads to a
configuration that either never was visited previously or was not visited for
some time greater than $k$, $k>0$ (if possible).

Among all possible evolutions, Markovian processes feature prominently since
they respect the fairness condition (it is well-known that Markovian processes
leave non-terminal strongly connected components with probability 1), they do
not require any history or global knowledge on the state space, and they
provide a modelling tool useful in many domains (like in the simulation of
population behaviours or in distributed algorithmics).
Such a Markovian process corresponds to a labeling of each state graph arrow
\Top{u}{u'}{p} by a static probability $p$ that only depends on configuration
$u$.
Here are two examples of such Markovian processes:
\begin{enumerate}
 \item \emph{Equiprobable evolutions}: \Top{u}{u'}{1/k_u} where $k_u$ denotes the
 cardinality of the set $\{u'\,|\,u\To{u'}\}$.
 \item \emph{Concentration-dependent evolutions}: \Top{u}{u'}{p_r} where $r$
 denotes the applied rule and $p_r$ is proportional to $h_r$, the number of
 distinct combinations of tokens that activate $r$, with a proportionality
 coefficient that only depends on $r$. Assuming that
 $r=\synt{q_1}{q_2}{q'_1}{q'_2}$, the number $h_r$ is given by
 \begin{equation}
  h_r=\left\{
       \begin{array}{ll}
	|u|_{q_1}|u|_{q_2} & \textnormal{if}\ q_1\ne 0,q_2\ne 0,q_1\ne q_2\\
	|u|_{q_1}(|u|_{q_1}-1) & \textnormal{if}\ q_1\ne 0,q_1=q_2\\
	|u|_{q_1} & \textnormal{if}\ q_2=0\\
	|u|_{q_2} & \textnormal{if}\ q_1=0
       \end{array}
      \right.
  \label{eq:hr}
 \end{equation}
       The two last cases hold when the environment (containing an infinite
       number of tokens) is involved in the rule.
\end{enumerate}

\subsection{FS-Mode GCPS modelling Population Dynamics}\label{ssec:dynamics}

Assuming that for a given rule proportionality coefficients are the same for all
configurations, the con\-cen\-tra\-tion-dependent strategy directly corresponds
to a run of the Gillespie's SSA. Thus, we advocate that GCPS in fs-mode provide
a good theoretical tool for studying population behaviours.

A paradigmatic example illustrating how GCPS allows a well suited specification
of population behaviours consists of the description of a process inspired by
the Lotka-Volterra model.

\paragraph*{The Lotka-Volterra Model.}
The Lotka-Volterra process was introduced by Lotka as a model of coupled
auto-catalytic chemical reactions, and was investigated by Volterra as a model
for studying an ecosystem of predators and preys~\cite{edelstein88}. This model
specifies how two coupled populations (of chemicals or individuals) $Y_1$ (the
preys) and $Y_2$ (the predators) behave. In~\cite{gillespie77}, D.T. Gillespie
proposes the study of this system derived from the following ODEs
\begin{equation}
\label{eq:ODElotkavolterra}
\begin{aligned}
\frac{dY_1}{dt}  & =  (c_1-c_2 Y_2)Y_1 & & & & & & & & 
\frac{dY_2}{dt} & =  (c_2 Y_1-c_3)Y_2
\end{aligned}
\end{equation}
Equivalently, the following chemical reactions
\begin{equation}
\label{eq:lotkavolterra}
\begin{aligned}
Y_1 & \mfd{c_1} 2\,Y_1 & & & & &
Y_1 + Y_2 & \mfd{c_2} 2\,Y_2 & & & & &
Y_2 & \mfd{c_3} .
\end{aligned}
\end{equation}
specify a model whose behaviour is described by ODEs
system~(\ref{eq:ODElotkavolterra}). The dynamics of these reactions is
conveniently characterized using the predator-prey interpretation. The first
rule states that a prey $Y_1$ reproduces. The second rule states that a predator
$Y_2$ reproduces after feeding on prey $Y_1$. Finally, the last rule specifies
that predators $Y_2$ die of natural causes. Coefficients $c_i$ are the rates of
the three reactions.
The correspondence between the two models relies in the fact that the trajectories
of the Gillespie's SSA tend to the solutions of the ODEs system given by the law
of mass action on the reactions. This result is due to the particular
application of the Kurtz's theorem~\cite{kurtz73} to chemical systems.

\begin{figure}[!p]
 \begin{center}
  \vspace{-10pt}
  \subfigure{\includegraphics[width=.49\textwidth]{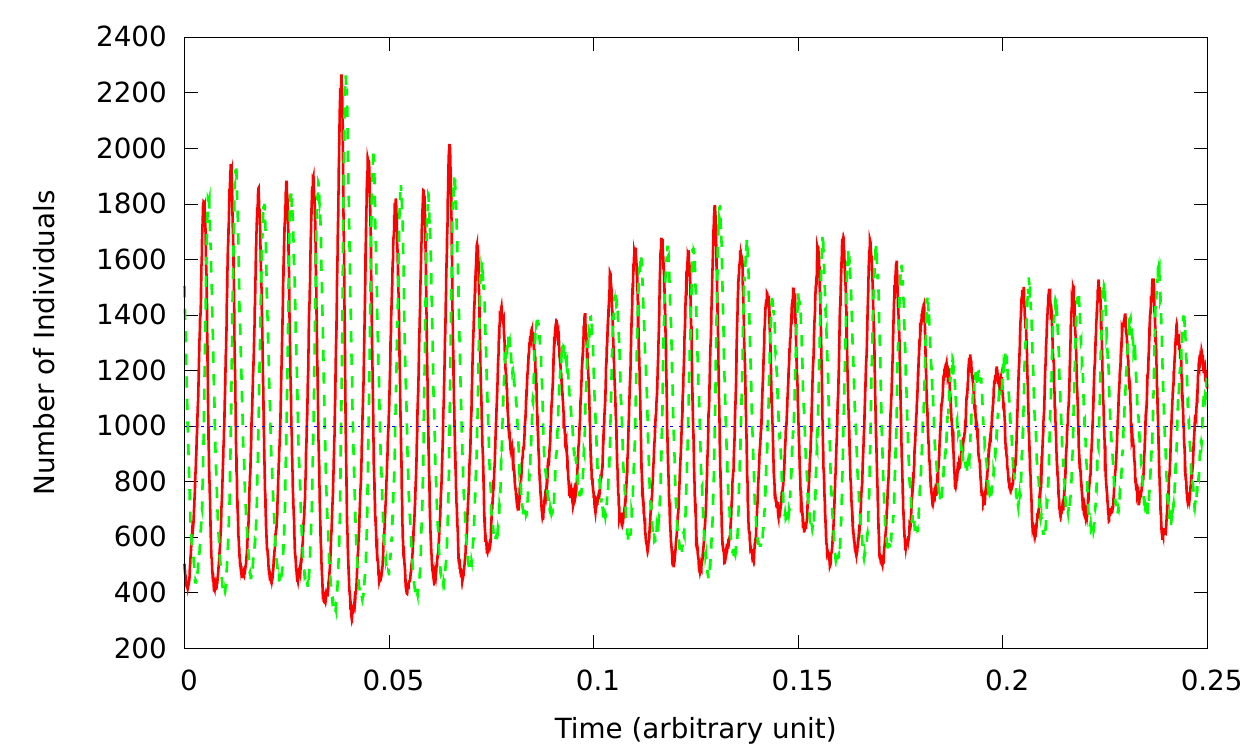}}
  \subfigure{\includegraphics[width=.49\textwidth]{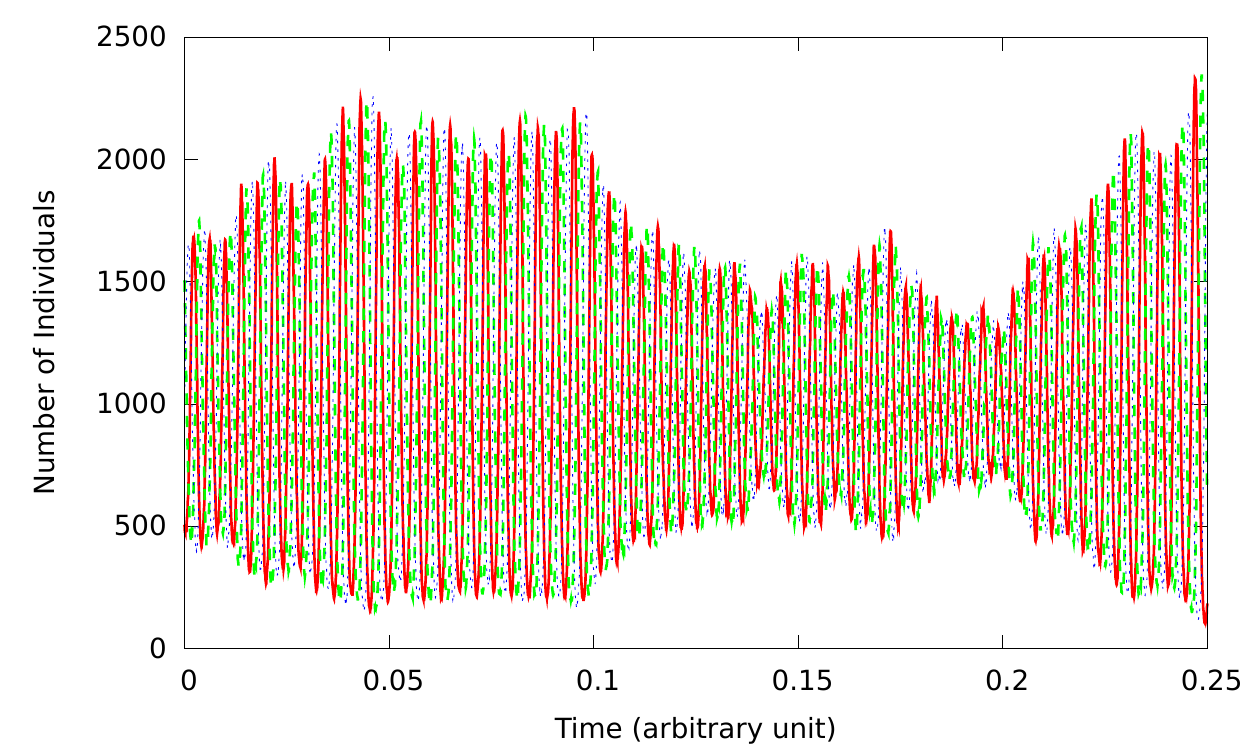}}
  \subfigure{\includegraphics[width=.49\textwidth]{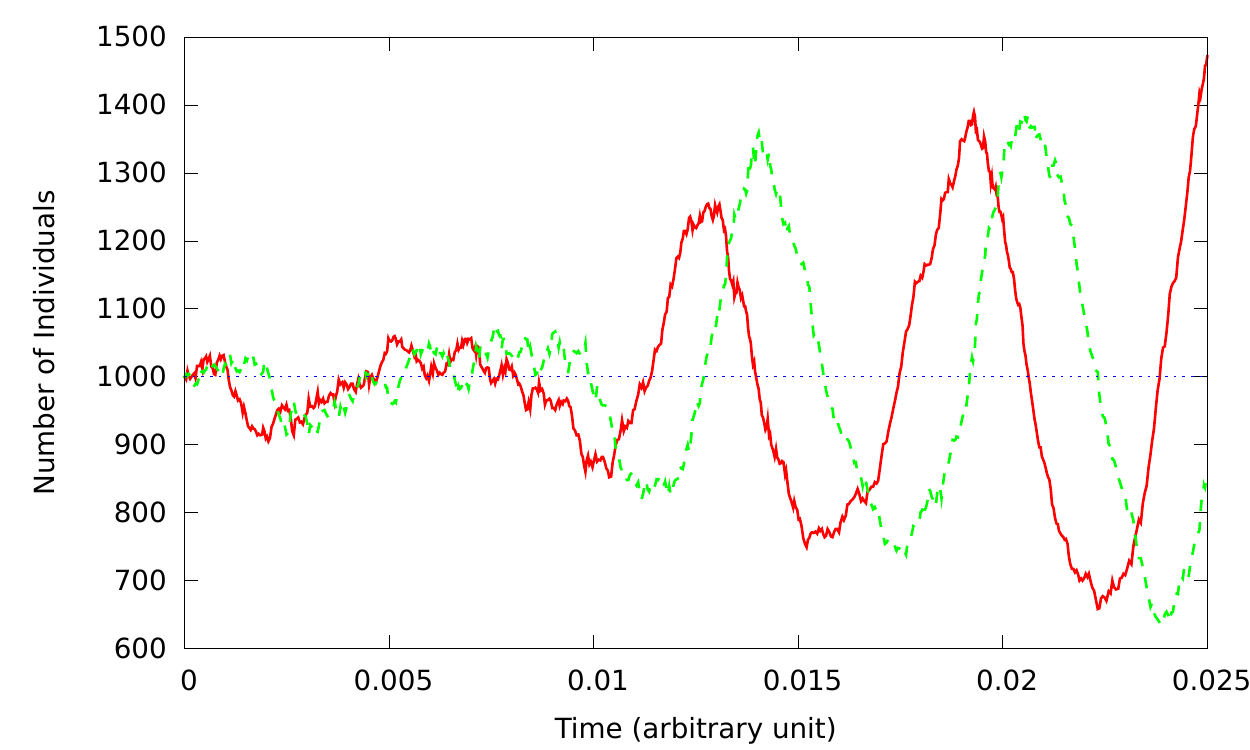}}
  \subfigure{\includegraphics[width=.49\textwidth]{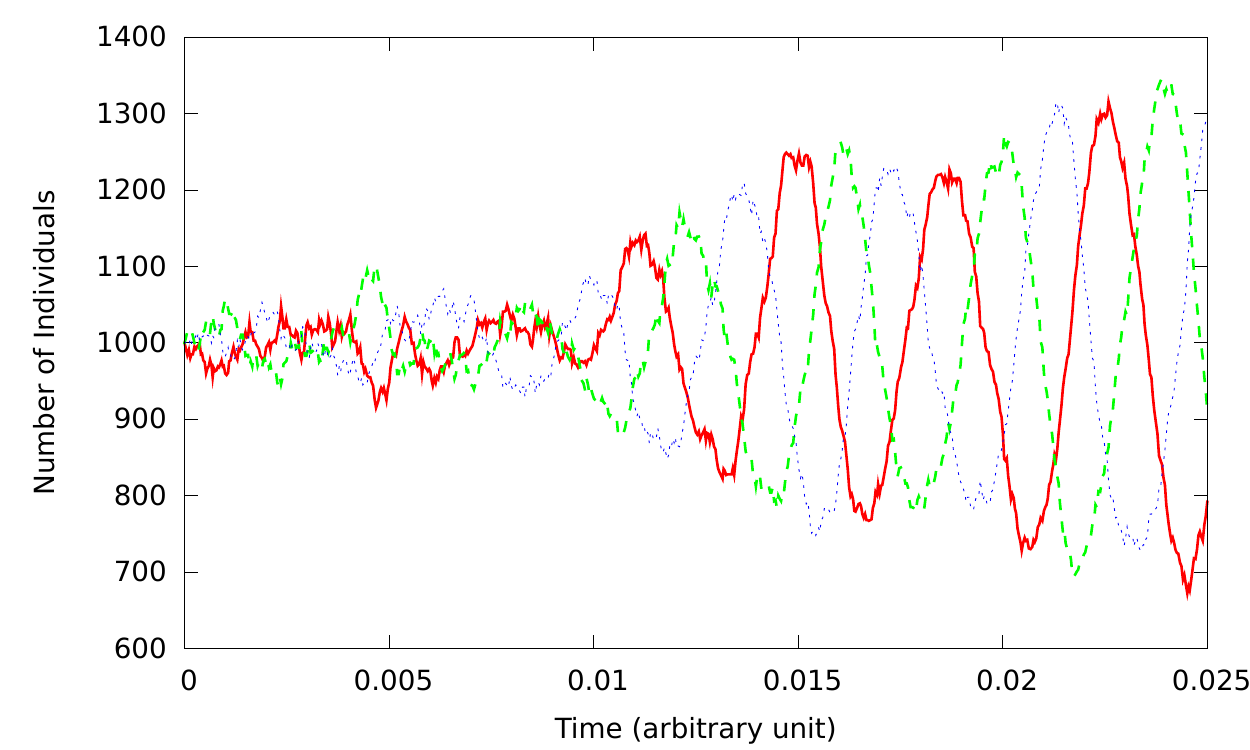}}
  \subfigure{\includegraphics[width=.49\textwidth]{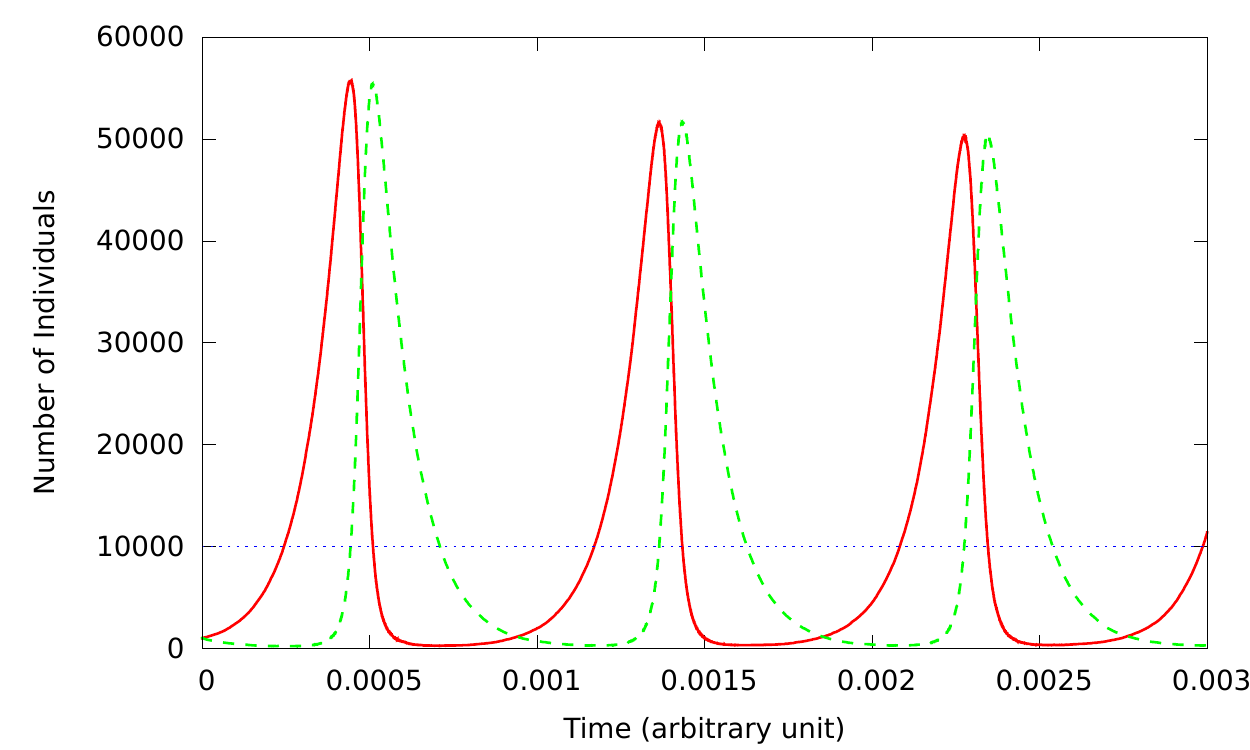}}
  \subfigure{\includegraphics[width=.49\textwidth]{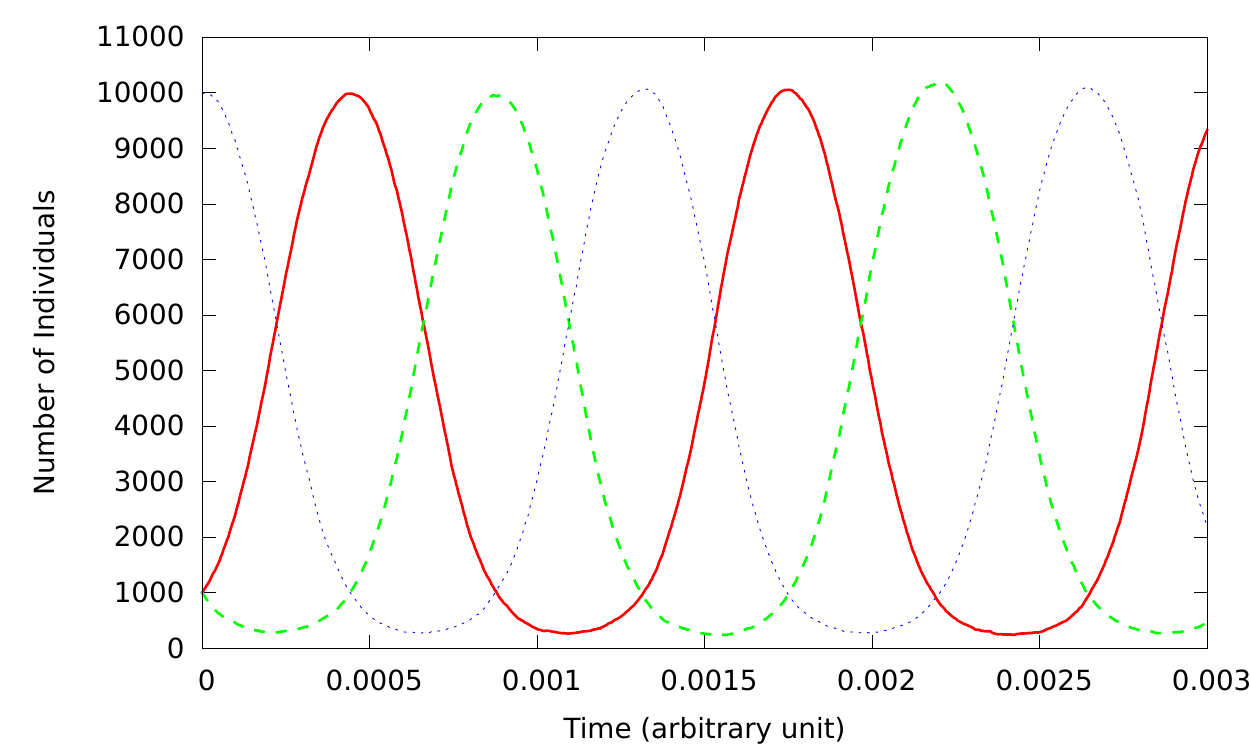}}
  \subfigure{\includegraphics[width=.49\textwidth]{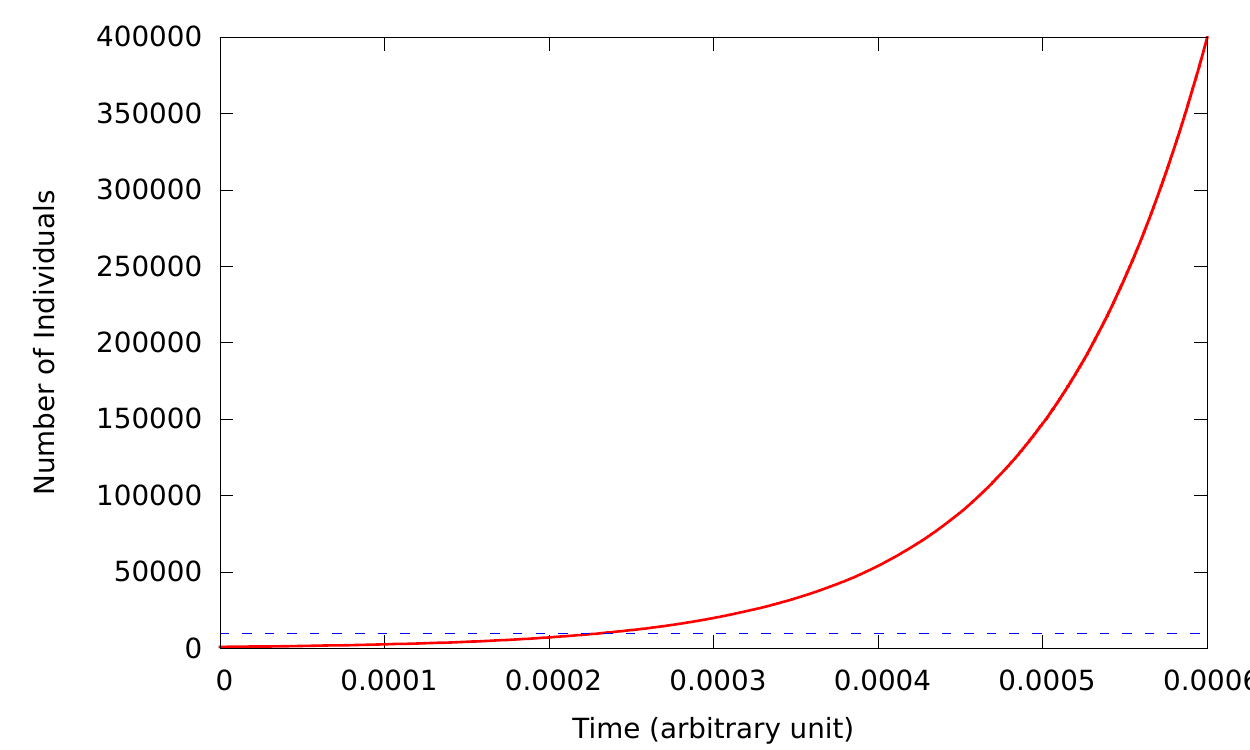}}
  \subfigure{\includegraphics[width=.49\textwidth]{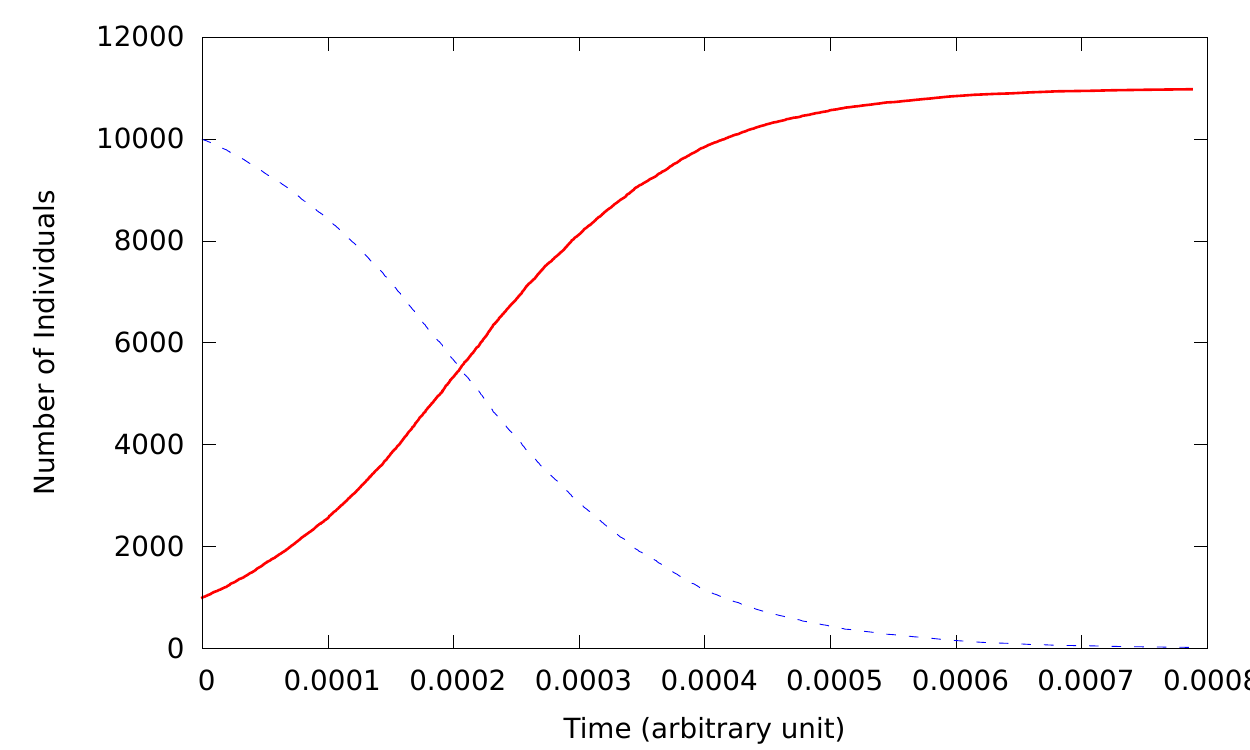}}
  \vspace{-5pt}
  \caption{\small Runs of the Lotka-Volterra model with renewable (left column)
  and not-renewable (right column) resources for different initial states
  (kinetics rates equal 1): $Y_1=500,Y_2=1500,\overline{X}=X=1000$ (first row),
  $Y_1=Y_2=\overline{X}=X=1000$ (second row),
  $Y_1=Y_2=1000,\overline{X}=X=10000$ (third row),
  $Y_1=1000,Y_2=0,\overline{X}=X=10000$ (fourth row). The solid red line
  represents preys, the dashed green line predators, and the blue dotted line
  resources. The two first rows show that both dynamics exhibit the same
  properties as presented in~\cite{gillespie77} (particularly, in second row,
  oscillations raise from an equilibrium initial state for the ODEs). The third
  row shows the difference in the dynamics when the resource size is ten times
  larger than the population size. The last row shows the difference in the
  dynamics when the predator population is empty. The simulations have been done
  using the general simulation language \mgs
  (\texttt{http://mgs.spatial-computing.org}) that allows an easy implementation
  of all models of the present
  article~\cite{spicher08,naco09,dubi10}.}\label{fig:lotkavolterra}
 \end{center}
\end{figure}

\paragraph*{Lotka-Volterra GCPS Definition.}
The model above does not fill GCPS requirements since the first and last
reactions are not pairwise interactions.
We propose to extend reactions~(\ref{eq:lotkavolterra}) by considering a
\emph{renewable} resource $\overline{X}$ for $Y_1$ as a third
species\footnote{We use the same notation as in~\cite{gillespie77} to express
that the food resource $X$ is assumed renewable.}: the molecular level of $X$
remains \emph{constant} whatever is its production or its consumption. The
extended system of reactions is:
\begin{equation}
\label{eq:GCPSlotkavolterra}
\begin{aligned}
\overline{X} + Y_1 \mfd{c'_1} 2\,Y_1 & & & & &
Y_1 + Y_2 \mfd{c'_2} 2\,Y_2 & & & & &
Y_2 + \overline{X} \mfd{c'_3} 2\,\overline{X}
\end{aligned}
\end{equation}
The use of a pairwise interaction in the last reaction can be interpreted as a
competition between the two predator behaviours: a predator in presence of preys
eates and reproduces (second reaction); a predator in absence of prey
(represented by the grass) dies (third reaction). Moreover, with the hypothesis
that the number of $\overline{X}$ remains constant, the behaviour of this system
is exactly described by ODEs~(\ref{eq:ODElotkavolterra}) with
$c_1=\overline{X}c'_1$, $c_2=c'_2$ and $c_3=\overline{X}c'_3$. Thus,
systems~(\ref{eq:lotkavolterra}) and~(\ref{eq:GCPSlotkavolterra}) are equivalent
in terms of dynamics.

System~(\ref{eq:GCPSlotkavolterra}) is only composed of pairwise interactions
that satisfy condition 4 of Definition~\ref{def:gencom}. Thus, it can be easily
translated to a one-symbol GCPS, denoted $\Pi_{LV}$, working in fs-mode (in
concentration-dependent implementation) with rules $R$:
\begin{equation*}
\begin{array}{l@{\qquad\qquad}l}
 \synt{0}{1}{1}{1}&\synt{1}{0}{1}{1}\\[2pt]
 \synt{1}{2}{2}{2}&\synt{2}{1}{2}{2}\\[2pt]
 \synt{2}{0}{0}{0}&\synt{0}{2}{0}{0}
\end{array}
\end{equation*}
where membrane indices $0$, $1$ and $2$ represent the environment (an infinite
source of $X$), the preys $Y_1$ and the predators $Y_2$, respectively.

Let now consider the previously defined concentration-dependent evolution with
probabilities $p_\mu=a_\mu/a_0$ for each $\mu\in{R}$ with the propensity
function $a_\mu=c_\mu\,h_\mu$: $c_\mu$ is the rate of the corresponding reaction
in~(\ref{eq:lotkavolterra}) and $h_\mu$ is given by equation~(\ref{eq:hr})
accordingly to $\Pi_{LV}$.
The reader is invited to pay attention that even if the environment is an
\emph{infinite} source of $X$ (instead of a \emph{constant} one), the
dynamics are well taken into account: rules involving the environment have
probabilities that do not depend on the environment size, see
equation~(\ref{eq:hr}). For example, the propensity of the first reaction is
given by $a_1=c_1\,h_1=c_1\,Y_1=c'_1\,\overline{X}\,Y_1$ as expected
w.r.t. reactions~(\ref{eq:GCPSlotkavolterra}).
In this respect, any computation of $\Pi_{LV}$ represents a run of the
Gillespie's SSA of reactions~(\ref{eq:GCPSlotkavolterra}). As a consequence,
$\Pi_{LV}$ is an exact model of the original Lotka-Volterra system.

It has to be remarked that $\Pi_{LV}$ cannot be described by any PP since the
environment objects are involved in its definition. A possible specification of
the Lotka-Volterra equations may be obtained within a PP by considering $X$ as a
\emph{not-renewable} resource. Such a definition has been realized (taking
reactions~(\ref{eq:GCPSlotkavolterra}) and substituting $\overline{X}$ by $X$.)
However, due to the limitation of resource, this system does not respect the
dynamics of equation~(\ref{eq:ODElotkavolterra}) anymore. For example, without
any predators, a population of preys stabilizes in this model, while in the
original model it grows exponentially.
Figure~\ref{fig:lotkavolterra} gives some examples of simulations of the
Lotka-Volterra model considering renewable and not-renewable resources.

\paragraph*{General Population Dynamics.}
It is possible to reverse the above method and to give a GCPS system whose
population dynamics will correspond to some dynamics given by a system of ODEs,
under the following conditions.
Let us consider the ODEs system defined on set of variables
$\{Y_1,\dots,Y_N\}$ of the form
\begin{equation}
 \frac{dY_i}{dt} = \sum_{j,k}a^i_{jk}Y_jY_k-\sum_j(b_{ij}+b_{ji})Y_iY_j\label{eq:generalODE}
\end{equation}
where coefficients $a^i_{jk}$ and $b_{ij}$ satisfy the following conditions:
\begin{enumerate}
 \item for all $i,j,k$, $a^i_{jk}\ge 0$ and $b_{ij}\ge 0$;
 \item for all $j,k$ such that $b_{jk}\ne 0$, there exists either one index $i_0$
 such that $a^{i_0}_{jk}=2\,b_{jk}$, or two distinct indices $i_1$ and $i_2$ such
 that $a^{i_1}_{jk}=a^{i_2}_{jk}=b_{jk}$; for any other index $i$, $i\ne{i_0}$
 or $i\ne{i_1}$ and $i\ne{i_2}$, $a^i_{jk}=0$.
\end{enumerate}
The above conditions are sufficient to ensure that $\sum_i\frac{dY_i}{dt}=0$.
Then there exists a concentration-dependent fs-mode GCPS without rules involving
the environment (\emph{i.e.}, a PP) whose behaviour is exactly described by
ODEs~(\ref{eq:generalODE}) when the population size goes to the
infinity. Indeed, these equations correspond to the mass-action law of a set of
rules such that for any $j,k$ with $b_{jk}\ne{0}$
\begin{equation*}
\syntp{j}{k}{i_0}{i_0}{b_{jk}}
\qquad\textnormal{or}\qquad
\syntp{j}{k}{i_1}{i_2}{b_{jk}}
\end{equation*}
according to the considered possibility of the above condition 2.
The reader is invited to pay attention that these equations correspond to a
wider range of dynamics than the dynamics of second-order chemical reactions
with two products since they allow the specification of ordered interactions
(e.g., involving a sender and a receiver as considered in the PP
literature). This property also holds in PP and suggests that
equations~(\ref{eq:generalODE}) exactly describe PP dynamics when the size of
the populations tends to the infinity.

It is obvious that a more general class of population behaviours is captured by
concentration-dependent fs-mode GCPS since they have not to be conservative
thanks to the environment. Following the idea of equivalence between
systems~(\ref{eq:lotkavolterra}) and~(\ref{eq:GCPSlotkavolterra}) in terms of
dynamics, equations~(\ref{eq:generalODE}) can be extended with the introduction
of a renewable variable $\overline{Y_0}$. This wider class of ODEs is supported
by concentration-dependent fs-mode GCPS model.

\section{Computational Properties}\label{sec:comp}

In this section, we focus on the original use of PP as a computational model of
algebraic numbers proposed in~\cite{Bournez}. This article investigates the
case where the computation is independent of the initial contents of the
system.

Using the GCPS terminology, the main idea of~\cite{Bournez} is to consider the
result of a computation as a ratio between the number of tokens in certain
membrane and the total number of tokens (without taking care of the environment)
when \emph{the population size goes to the infinity} and when \emph{the state of the
system converges}.
The proposed work relies on the definition of a particular strategy of execution
of the PP: a step of execution consists in sampling uniformly and independently
of the past two distinct tokens in the membrane and let them interact in a
sequential mode. This strategy is fair since it corresponds to a Markov process.
The authors of the aforementioned article studied the Markov chain associated
with PP and proved its equivalence to some system of ODEs at the limit.

We remark that the same kind of result directly arises from considerations of
Section~\ref{ssec:dynamics} since this computational model is captured by
one-symbol GCPS working in fs-mode with Gillespie concentration-dependent
implementation. Indeed, the above execution strategy exactly corresponds to a
Gillespie's SSA run where the stochastic constants equal 1 for all rules. Thus,
the study of the model corresponds to the investigation of the sensibility of
the associated ODEs system.
Let us illustrate this point by considering the running example
of~\cite{Bournez}
\begin{eqnarray*}
 \synt{p}{p}{p}{m}\\[2pt]
 \synt{p}{m}{p}{p}\\[2pt]
 \synt{m}{p}{p}{p}\\[2pt]
 \synt{m}{m}{p}{m}
\end{eqnarray*}
where symbols $p$ and $m$ identify two membranes. It has been shown that the
ratio $\frac{p}{p+m}$, where $p$ (resp.~$m$) is the size of the membrane $p$
(resp. $m$), converges to $\frac{1}{\sqrt{2}}$ when the population size goes to
the infinity.
Accordingly to equations~(\ref{eq:generalODE}), we associate ODEs with this GCPS
as follows
\begin{equation*}
\begin{aligned}
\frac{dY_p}{dt} & = Y_m^2+2\,Y_pY_m-Y_p^2 & & & & & & & & &
\frac{dY_m}{dt} & = -Y_m^2-2\,Y_pY_m+Y_p^2
\end{aligned}
\end{equation*}
The stable states of this system are obtained when the two equations vanish,
that is, when either $Y_m=-(\sqrt{2}+1)Y_p$ or $Y_m=(\sqrt{2}-1)Y_p$. The first
solution is incoherent since it involves a negative size of population. The
second solution trivially leads to the expected result
$\frac{Y_p}{Y_p+Y_m}=\frac{1}{\sqrt{2}}$.

\section{Conclusions}

In this article we investigated connections between population protocols and
generalized communicating P systems. The two models share the same multiset
structure and the same type of rules. Traditionally PP are used to study
population dynamics in the context of distributed algorithmics while GCPS are
investigated for the computational properties.

By incorporating the derivation mode from PP into GCPS framework we obtained a
strict inclusion of PP in GCPS working in fs-mode. We then took a particular
implementation of the fs-mode corresponding to a run of the Gillespie's SSA and
we obtained that the dynamics of the systems can be described by the
corresponding system of differential equations. Different questions then could
be explored, like the investigation of the conditions ensuring that the system
reaches a stable state regardless of its initial state or ensuring that a stable
state is never reached for any initial configuration. GCPS are in this sense
easier to handle than PP because of the environment that permits to easily
simulate the equivalent of creation or degradation reactions.
Section~\ref{ssec:dynamics} also considers the converse problem of the
construction of a GCPS system exhibiting a particular behaviour given by a
systems of ODEs. It would be interesting to see if the given sufficient
conditions are also necessary.  A mathematical challenge resulting from
Section~\ref{sec:comp} is whether for any algebraic number $x\in[0..1]$ there is
a GCPS working in concentration-dependent evolution implementation of the
fs-mode that converges to $x$.

We remark that the presented results hold only in the concentration-de\-pen\-dent
implementation of the fs-mode. By taking an equiprobable implementation the
results are completely different.

Since Petri Nets can be seen as multiset rewriting, it is clear that the
results of this paper can be translated to this domain (for Petri Nets with
specific type of rules and an additional fairness strategy).

We think that the fs-mode has interesting properties that should be further
explored. As showed in the article,  the fairness condition is in some sense
similar to a stochastic evolution, so it could be preferable to consider this
condition instead of a stochastic behaviour. Another interesting property of the
proposed stochastic implementation is that Gillespie's SSA introduces an
explicit continuous time and discrete events in the model, which do not appear
in a GCPS description.

\paragraph*{Acknowledgments.}
The authors would like to acknowledge the support of ANR project SynBioTIC.

\bibliographystyle{abbrv}

\begin{thebibliography}{10}

\bibitem{AADFP04}
D.~Angluin, J.~Aspnes, Z.~Diamadi, M.~J. Fischer, and R.~Peralta.
\newblock Computation in Networks of Passively Mobile Finite-State Sensors.
\newblock {\em Distributed Computing}, pages 235--253, Mar. 2006.

\bibitem{AAE06}
D.~Angluin, J.~Aspnes, and D.~Eisenstat.
\newblock Stably Computable Predicates are Semilinear.
\newblock In {\em PODC'06: Proceedings of the twenty-fifth annual ACM symposium
  on Principles of distributed computing}, pages 292--299, New York, NY, USA,
  2006. ACM Press.

\bibitem{AR07}
J.~Aspnes and E.~Ruppert.
\newblock An Introduction to Population Protocols.
\newblock {\em Bulletin of the Europ. Assoc. for Theor. Comp. Sci.},
  93:98--117, Oct. 2007.

\bibitem{BGMV07}
F.~Bernardini, M.~Gheorghe, M.~Margenstern, and S.~Verlan.
\newblock How to Synchronize the Activity of All Components of a {P} System?
\newblock {\em International Journal of Foundations of Computer Science.},
  19(5):1183--1198, 2008.

\bibitem{Bournez}
O.~Bournez, P.~Chassaing, J.~Cohen, L.~Gerin, and X.~Koegler.
\newblock On the Convergence of Population Protocols when Population Goes to
  Infinity.
\newblock {\em Applied Mathematics and Computation}, 215(4):1340--1350, 2009.

\bibitem{minpar}
G.~Ciobanu, L.~Pan, G.~Paun, and M.~J. P{\'e}rez-Jim{\'e}nez.
\newblock P Systems with Minimal Parallelism.
\newblock {\em Theor. Comput. Sci.}, 378(1):117--130, 2007.

\bibitem{CVV10}
E.~Csuhaj-Varj\'{u}, G.~Vaszil, and S.~Verlan.
\newblock On Generalized Communicating {P} Systems with One Symbol.
\newblock In M.~Gheorghe, T.~Hinze, and G.~Paun, editors, {\em Proceedings of
  the Eleventh International Conference on Membrane Computing}, pages 137--154.
  Verlag ProBusiness Berlin, 2010.

\bibitem{CV11}
E.~Csuhaj-Varj{\'u} and S.~Verlan.
\newblock On Generalized Communicating {P} Systems with Minimal Interaction
  Rules.
\newblock {\em Theor. Comp. Sci.}, 412(1-2):124--135, 2011.

\bibitem{edelstein88}
L.~Edelstein-Keshet.
\newblock {\em Mathematical Models in Biology}.
\newblock Random House, New York, 1988.

\bibitem{FreundVerlan07}
R.~Freund and S.~Verlan.
\newblock A Formal Framework for Static (Tissue) {P} Systems.
\newblock In G.~Eleftherakis, P.~Kefalas, G.~P\u{a}un, G.~Rozenberg, and
  A.~Salomaa, editors, {\em Membrane Computing, 8th International Workshop, WMC
  2007, Thessaloniki, Greece, June 25-28, 2007 Revised Selected and Invited
  Papers}, volume 4860 of {\em LNCS}, pages 271--284. Springer, 2007.

\bibitem{friscoBook}
P.~Frisco.
\newblock {\em Computing with Cells}.
\newblock Oxford University Press, 2009.

\bibitem{gillespie77}
D.~T. Gillespie.
\newblock Exact Stochastic Simulation of Coupled Chemical Reactions.
\newblock {\em J. Phys. Chem.}, 81(25):2340--2361, 1977.

\bibitem{kurtz73}
T.~Kurtz.
\newblock A Limit Theorem for Perturbed Operator Semigroups with Applications
  to Random Evolutions.
\newblock {\em J. Funct. Anal.}, 12:55--67, 1973.

\bibitem{naco09}
O.~Michel, A.~Spicher, and J.-L. Giavitto.
\newblock Rule-Based Programming for Integrative Biological modelling --
  Application to the modelling of the Lambda Phage Genetic Switch.
\newblock {\em Natural Computing}, 8(4):865--889, december 2009.

\bibitem{Presburger29}
M.~Presburger.
\newblock Uber die Vollstandig-keit eines Gewissen Systems der Arithmetik
  Ganzer Zahlen, in Welchemdie Addition als Einzige Operation Hervortritt.
\newblock {\em Comptes rendus du I Congres des Mathematicians des Pays Slaves},
  pages 92--101, 1929.

\bibitem{Pbook}
G.~P\u{a}un.
\newblock {\em Membrane Computing. An Introduction}.
\newblock Springer--Verlag, 2002.

\bibitem{Phandbook}
G.~P\u{a}un, G.~Rozenberg, and A.~Salomaa.
\newblock {\em The Oxford Handbook Of Membrane Computing}.
\newblock Oxford University Press, 2009.

\bibitem{PN}
W.~Reisig.
\newblock {\em Petri {N}ets. An Introduction}.
\newblock Springer, 1985.

\bibitem{HFL}
G.~Rozenberg and A.~Salomaa.
\newblock {\em Handbook of Formal Languages, 3 volumes}.
\newblock Springer, 1997.

\bibitem{spicher08}
A.~Spicher, O.~Michel, M.~Cieslak, J.-L. Giavitto, and P.~Prusinkiewicz.
\newblock Stochastic P Systems and the Simulation of Biochemical Processes with
  Dynamic Compartments.
\newblock {\em BioSystems}, 91(3):458--472, March 2008.

\bibitem{dubi10}
A.~Spicher, O.~Michel, and J.-L. Giavitto.
\newblock {\em Understanding the Dynamics of Biological Systems}, chapter
  Interaction-Based Simulations for Integrative Spatial Systems Biology, 195--231.
\newblock Springer, 2011.

\bibitem{VBGM08}
S.~Verlan, F.~Bernardini, M.~Gheorghe, and M.~Margenstern.
\newblock Generalized Communicating {P} Systems.
\newblock {\em Theor. Comp. Sci.}, 404(1-2):170--184, 2008.

\end{thebibliography}

\end{document}